# A comparative study of divisive hierarchical clustering algorithms

Maurice Roux

`mrhroux@yahoo.fr`

ABSTRACT. A general scheme for divisive hierarchical clustering algorithms is proposed. It is made of three main steps : first a splitting procedure for the subdivision of clusters into two subclusters, second a local evaluation of the bipartitions resulting from the tentative splits and, third, a formula for determining the nodes levels of the resulting dendrogram.

A number of such algorithms is presented. These algorithms are compared using the Goodman-Kruskal correlation coefficient. As a global criterion it is an internal goodness-of-fit measure based on the set order induced by the hierarchy compared to the order associated with the given dissimilarities.

Applied to a hundred random data tables, these comparisons are in favor of two methods based on unusual ratio-type formulas for the splitting procedures, namely the Silhouette criterion and Dunn's criterion. These two criteria take into account both the within cluster and the between cluster mean dissimilarity. In general the results of these two algorithms are better than the classical Agglomerative Average Link method.

## 1  Introduction

Most papers relative to hierarchical clusterings use one of the four popular agglomerative methods, namely the single link method, the average link method, the complete link method and Ward's method. The goal of these methods is to represent the proximities, or the dissimilarities, between objects as a tree where the objects are situated at the end of the branches, generally at the bottom of the graph (Fig. 1). The junctions of the branches are called the nodes of the tree; the node levels are supposed to represent the intensity of the ressemblance between the objects or clusters being joined.

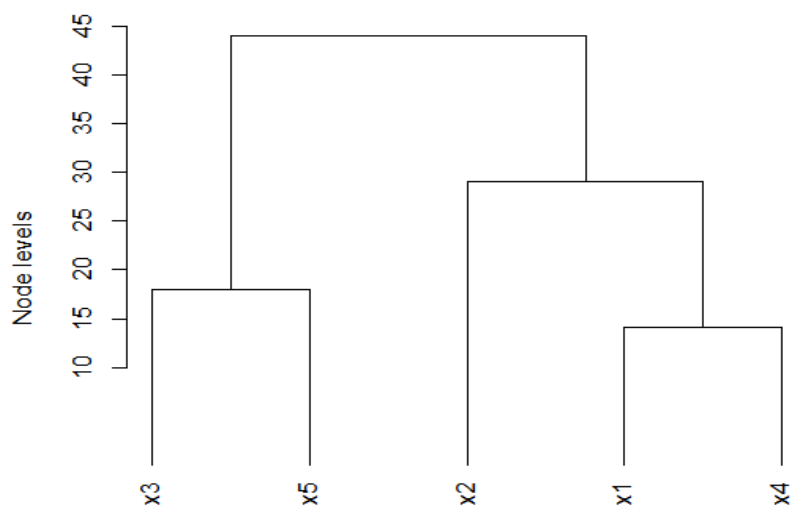

Figure 1. Dendrogram resulting from a hierachical clustering program

In an agglomerative procedure (coined SAHN for Sequential Agglomerative Hierarchical Non-overlapping by Sneath and Sokal 1973) the tree is constructed bottom-up: at the beginning each object $x$ is considered as a cluster $\{x\}$ called a singleton. Then each step of the procedure consists in creating a new cluster by merging the two closest clusters. This implies there is a way to compute the dissimilarity or distance between two clusters. For instance, in the usual average link method the distance between two clusters $C_p$ and $C_q$ is the mean value of the between-cluster distances:

$$D(C_p, C_q) = (1 / n_p n_q) \Sigma \{ d(x_i, x_j) \mid x_i \in C_p, x_j \in C_q \} \quad (1)$$

where $n_p$ and $n_q$ are the number of elements of $C_p$ and $C_q$ respectively, $d(x_i, x_j)$ is the given dissimilarity, or distance, between objects $x_i$ and $x_j$.

The value of $D(C_p, C_q)$ is then used as the node level for the junction of the branches issued from $C_p$ and $C_q$. Indeed it can be shown that, this way, the usual procedures are monotonic. This means that if cluster C is included in a cluster C′ then their associated node levels $L_C$ and $L_{C'}$ are in an increasing order:

$$C \subseteq C' \Rightarrow L_C \leq L_{C'} \quad (2)$$

This ensures that the hierarchical tree may be built without branch crossings. Thus formula (1) is used first as a criterion for merging the clusters and, second, for determining the node levels of the hierarchy.

Divisive hierachical algorithms are built top-down: starting with the whole sample in a unique cluster they split this cluster into two subclusters which are, in turn, divided into subclusters and so on. At each step the two new clusters make up a so-called bipartition of the former. It is well known (Edwards and Cavalli-Sforza 1965) that there are $2^{n-1} - 1$ ways of splitting a set of n objects into two subsets. Therefore it is too time consuming to base a splitting protocol on the trial of all possible bipartitions. The present paper proposes to evaluate a restricted number of bipartitions to make up a working algorithm. Such an idea was developed a long time ago by Macnaughton-Smith *et al.* (1964) and reused by Kaufman & Rousseeuw (1990).

The main objective of this study is to propose a general scheme for the elaboration of divisive hierarchical algorithms, where three main choices should apply at each step of the procedure:

i) a simplified way of splitting the clusters
ii) a formula to evaluate each of the considered bipartitions
iii) a formula to determine the node levels of the resulting hierarchy

The present work aims at the treatment of small to moderate size datasets (a hundred objects or so) but with a search for the quality of the results. In this framework only complete binary hierarchies are looked for, so the choice of which cluster to split is not relevant: all clusters including two or more objects are split in turn, until there remain only singletons.

The above three points will be studied in the following (sections 2, 3 and 4). Applying these principles gives rise to a family of algorithms described in section 5. Then a practical benchtest is made up and used (section 6) for comparing the new algorithms with the classical average link procedure. The main results are gathered (section 6.4) and a concluding section terminates this paper (section 7).

## 2 Splitting procedures

A number of splitting procedures were designed in the past, the oldest one being by Williams & Lambert (1959). This procedure is said to be monothetic in the sense that object sets are split according to the values of only one variable. This idea has been updated using one principal component instead of a single variable (algorithm Principal Directions Divisive Partitioning or PDDP by Boley 1997).

Another approach to get around the complexity of splitting is to extract one, or several objects, from the set to be split. Macnaughton-Smith et al.(1964) proposed to select the most distant object from the cluster as a seed for a separate new cluster. Then they aggregate to this seed the objects which are closer to the new subset than to the rest of the current cluster (see section 5.3 for details). A similar idea was developed by Hubert (1973): he suggested to use a pair of objects as seeds for the new bipartition. His choice was to select the two objects that are most dissimilar, and then to build up the two subclusters according to distances (or a function of distances) to these seeds. Exploiting this idea Roux (1991, 1995) considered the bipartitions generated by all the pairs of objects, retaining the bipartition with the best evaluation of some a priori criterion. This procedure will be applied in the following.

Finally another divisive algorithm was based on the usual k-means method for partitioning a set of objects. Called the Bisecting k-means (Steinbach et al. 2000) this procedure builds up the successive dichotomies by a 2-means algorithm with either a random initial partition or with a partition using one of the above procedures.

## 3 Evaluation of bipartitions

At each step of a usual agglomerative method the two candidate clusters $C_p$ and $C_q$ for a merging step may be considered as the bipartition { $C_p$ , $C_q$ } of the set $C_p \cup C_q$ . For instance, in the case of the classical average link method, such a bipartition is evaluated by the mean value of the between-cluster distances.

In divisive methods, once the cluster $C_p$ to be split is selected, the next step is to study a number of bipartitions {$C'_p$ , $C''_p$ } of $C_p$ . Again the between-cluster average distances can be used for evaluating this split (Roux 1991). Another criterion was suggested in an agglomerative framework (Mollineda and Vidal 2000). Moreover a number of criteria designed for the evaluation of any partition can be used. Some of them are described in this section which will be used in the applications (section 6).

Whatever the adopted criterion, it should be noted that a series of very good bipartitions does not result automatically in a good hierarchy.

### 3.1 Distance-like criteria

The single link criterion is the shortest distance between objects of $C_p$ and objects of $C_q$ :

$$D_{SL}(C_p , C_q ) = \text{Min} \{ d(x_i, x_j) \mid x_i \in C_p , x_j \in C_q \}$$

Similarly the complete link criterion is the largest distance between objects of $C_p$ and objects of $C_q$ :

$$D_{CL}(C_p, C_q) = \text{Max} \{ d(x_i, x_j) \mid x_i \in C_p, x_j \in C_q \}$$

The average link criterion is defined by formula (1).

Ward's criterion may also be considered as a distance-like criterion. Indeed, after the paper of Székely & Rizzo (2005), there exists an infinite family of algorithms similar to Ward's. In the present study the focus is only on two of them. One is the original algorithm as described by J.H. Ward (1963). The second is defined by the parameter $\alpha = 1$ in the Székely-Rizzo family. Here under the between-cluster distances involved by these algorithms are designated as $D_{W1}$ (Ward's original) and $D_{W2}$ respectively, after Murtagh & Legendre (2014).

$$D_{W1}(C_p, C_q) = (n_p n_q / n_p + n_q) [ (2 / n_p n_q) \Sigma \{ d^2(x_i, x_j) \mid x_i \in C_p, x_j \in C_q \} \ldots$$
$$- (1 / n^2_p) \Sigma \{ d^2(x_i, x_j) \mid x_i \in C_p, x_j \in C_p \} - (1 / n^2_q) \Sigma \{ d^2(x_i, x_j) \mid x_i \in C_q, x_j \in C_q \} ]$$

If the objects $x$ are embedded in a vector space of real numbers then this formula may be rewritten as:

$$D_{W1}(C_p, C_q) = (n_p n_q / n_p + n_q) \| \bar{x}_p - \bar{x}_q \|^2$$

where $\bar{x}_p$ and $\bar{x}_q$ are the centroids of clusters $C_p$ and $C_q$ respectively. This formula shows that, in this case, $D_{W1}$ is proportional to the between centroids squared distance.

$$D_{W2}(C_p, C_q) = (n_p n_q / n_p + n_q) [ (2 / n_p n_q) \Sigma \{ d(x_i, x_j) \mid x_i \in C_p, x_j \in C_q \} \ldots$$
$$- (1 / n^2_p) \Sigma \{ d(x_i, x_j) \mid x_i \in C_p, x_j \in C_p \} - (1 / n^2_q) \Sigma \{ d(x_i, x_j) \mid x_i \in C_q, x_j \in C_q \} ]$$

This second formula is very similar to the first one except for the exponent on the initial distances $d(x_i, x_j)$.

### 3.2 Ratio-type criteria

The main idea for using such criteria is to take into account, not only the betwen cluster distances, but also the distances to the neighboring objects of the two clusters being studied. A popular, though ancient, criterion is that of Dunn (1974), designed to evaluate a partition P with any number of elements:

$$\text{Dunn}(P) = \text{Min}\{ \text{Min} \{ \bar{d}(C_p, C_q) \mid q \in P, q \neq p \}, p \in P \} / \text{Max} \{ \Delta_p \mid p \in P \}$$

where $\bar{d}(C_p, C_q)$ is the mean value of between-cluster distances and $\Delta_p$ is the diameter of subset $C_p$ (i.e. the largest distance between objects included in $C_p$).

When evaluating a bipartition made of C′ and C″ the above formula reduces to:

$$D_{Du}(C', C'') = \bar{d}(C', C'') / \text{Max}\{\Delta', \Delta''\}$$

But it is known that the diameter is rather sensitive to possible outliers; this is why the following variant is taken into account in the present study:

$$D_{Du}(C', C'') = \bar{d}(C', C'') / \text{Max}\{ \bar{d}(C'), \bar{d}(C'')\}$$

where $\bar{d}(C')$ (resp. $\bar{d}(C'')$) is the average value of within-cluster $C'$ (resp. $C''$) distances.

Finally the last criterion considered in this work is the Silhouette width (Kaufman & Rousseeuw, 1990). For any object $x_i$, included in a cluster $C(x_i)$, two functions, a and b, are defined and combined to get the silhouette $s(x_i)$ of this object:

$$a(x_i) = (\ 1\ /\ (\text{Card}(C(x_i)) - 1)\ \Sigma\ \{\ d(x_i, x_j)\ |\ x_j \in C(x_i)\ \} = \bar{d}(\{x_i\}, C(x_i) - \{x_i\})$$

$$b(x_i) = \text{Min}\{\ \bar{d}(\{x_i\}, C_p)\ |\ C_p \in P - C(x_i)\ \}$$

$$s(x_i) = (\ b(x_i) - a(x_i))\ /\ \text{Max}\ \{\ a(x_i), b(x_i)\ \}$$

The Silhouette width $S(P)$ of a partition P is just the mean value of all the $s(x_i)$ for the $x_i$ covered by P

$$S(P) = (1\ /\ n)\ \Sigma\ \{\ s(x_i)\ |\ x_i \in \cup\ \{C_p\ |\ p \in P\}\}$$

with n being the number of objects concerned by the current partition P. When C includes a bipartition $\{C', C''\}$ the functions a and b become:

for $x_i$ in $C'$    $a(x_i) = \bar{d}(\{x_i\}, C' - \{x_i\})$    $b(x_i) = \bar{d}(\{x_i\}, C'')$

for $x_i$ in $C''$   $a(x_i) = \bar{d}(\{x_i\}, C'' - \{x_i\})$   $b(x_i) = \bar{d}(\{x_i\}, C')$

while the formal definitions of $s(x_i)$ and $S(C)$ remain unchanged.

## 4   Determining the node levels

When used in an agglomerative scheme the usual five criteria examined in Section 3.1 are used without any problem for the representation of the results: the criterion value becomes the level of the corresponding node, and the drawing of the hierarchical tree does not show any cross-over (or reversal) of the branches.

Unfortunately divisive procedures, in general, do not enjoy this property, because of the non-optimality of the successive splittings. A rule is then needed to obtain consistent node levels and a true tree representation. Kaufman & Rousseeuw (1990), in their program DIANA, use the diameter of the successive clusters as node levels. It is evident that the diameter of a subset $C'$ included in a set C is less than, or equal to the diameter of C, fulfilling the monotonic condition (2). Thus the two subsets created by the splitting of C are always associated with lower (or equal) node levels.

Another way to settle consistent node levels would be to associate the ranks of the nodes according to the order in which they are created, starting with rank n – 1 for the top level, down to 1 for the last created node. But this method may not be satisfying; in effect it may happen that a small homogeneous susbet would be separated at an early stage from the bulk of the objects. The corresponding node would then be associated with a high rank, in spite of its homogeneity. Another way to use ranks would be to renumber the nodes from the bottom up to the top, after completion of all the splittings, but this is not free of difficulties either.

Indeed the present discussion is of little use for the general purpose of comparing clustering algorithms, since the global evaluation of the results will be based on rank correlation methods (see Section 6.2). However the users could be interested in getting coherent node levels, hence a working representation. In the present experiment the node levels are determined, as in the program DIANA, by their diameters.

## 5  New algorithms and an old one

A set of ten divisive algorithms is studied, though this set does not exhaust the possibilities of other algorithms following the principles of section 1. In the following list (Table 1) they are enumerated with the type of their bipartition criteria. As a basic reference the usual Average Link agglomerative method is included in this study.

### 5.1  Distance-like criteria

In the present work the divisive algorithms are based on the two-seeds splitting procedure (Section 2) except in the Macnaughton-Smith method and in the PDDP algorithm. Contrary to the rule applied in agglomerative methods, the highest value of the criterion indicates, in a divisive scheme, the bipartion to use. In case of ties the first appeared split is selected. However a small modification is adopted in the "Complete Link Divisive" method: the splitting criterion is to the smallest values for the diameters of the two candidate subsets, instead of the highest between-cluster distance.

|  | Type of criterion |
|---:|:---:|
| **Single link** | Dist. |
| **Average link divisive** | Dist. |
| **Complete link** | Dist. |
| **Ward's original** | Dist. |
| **Ward's Szekely-Rizzo** | Dist. |
| **Dunn's original** | Ratio |
| **Dunn's variant** | Ratio |
| **Silhouette** | Ratio |
| **Principal direction (PDDP)** | ? |
| **Macnaughton-Smith** *et al.* | ? |
| **Average link agglomerative** | Dist. |

Table 1. List of the 11 divisive procedures with their criteria.

### 5.2  Ratio-type criteria

The two variants of Dunn's formula are used and the Silhouette formula (Section 3.2) is also applied. As for distance-like formulas the successive splits with ratio-type criteria are those which maximize these criteria.

### 5.3  Three more algorithms

For the sake of comparisons two existing algorithms are added to the above list together with the classical Average Link Agglomerative method. The first one is the Principal Directions Divisive Partitioning (PDDP, Boley, 1998) and the other second one is the algorithm proposed by Macnaughton-Smith *et al.* (1964); they are listed at the end of Table 1. The

PDDP method was first designed for the analysis of observations × variables data tables, but may be readily adapted by using the Principal Coordinates Analysis (PCoA, Gower 1966). This technique is akin to the Principal components analysis (PCA). In the PDDP algorithm the first principal coordinate axis is used to create a dichotomy: those objects whose coordinates are negative are put into the first subset of the dichotomy, while the objects with positive, or null, coordinates make up the second subset. However a further step is taken which may move some objects from their actual assignment to the other one as long as they are closer to the latter than to the former. This PCoA is recomputed for each cluster with more than two objects, achieving a hierarchical divisive procedure.

The method proposed by Macnaughton-Smith *et al.* could be considered as a one-seed procedure. To split the cluster C they choose as seed the object $x_0$ whose average distance to the other elements of C is maximum. The building of the bipartition begins with

$$C' = C - \{x_0\} \text{ and } C'' = \{x_0\}$$

Next, for each object $x$ in C′, compute $\bar{d}(\{x\}, C' - \{x\})$ and $\bar{d}(\{x\}, C'')$, and retain $x_1$ as the one which maximizes

$$f(x) = \bar{d}(\{x\}, C'') - \bar{d}(\{x\}, C' - \{x\})$$

Then the bipartition becomes

$$C'_1 = C' - \{x_1\} \text{ and } C''_1 = C'' \cup \{x_1\}$$

this process is continued until $f(x)$ becomes negative.

Finally the usual Average Link Agglomerative method is added as a basic ground for the comparisons.

## 6 Practical tests

The comparison of the above 11 algorithms is restricted to the quality of the results as measured by the goodness-of-fit of the results to the data. First the benchmark made of random data sets is described, second some thoughts about the goodness-of-fit criteria are developed. Third a tentative estimation of the algorithmic complexity is studied and, finally, a summary of the comparisons is set up.

### 6.1 Random data sets

A sample of 100 rectangular 40 × 10 matrices are generated from a uniform distribution over [0, 1]. All 10 variables are generated independently according to the same distribution. For each matrix the usual Euclidean distance is applied in order to obtain a set of 100 distance matrices as input in the clustering programs. Although these data are far from real life data they constitute a harsh benchmark and allow for a real competition among the programs.

### 6.2 A goodness-of-fit criterion

The most popular criterion to evaluate the results of hierarchical clusterings is certainly the Co-phenetic Correlation Coefficient (CPCC, Sokal & Rohlf, 1962). It needs the construction of the ultrametric distances associated with the dendrogram; then the CPCC is just the usual correlation coefficient between the input distances and the ultrametric distances, the values of which are laid out in two long vectors.

But in the present work the focus is rather oriented toward rank correlation methods. Indeed, when the user examines the hierarchy issued from the data, the user focuses mainly on the groups, and subgroups, disclosed by the algorithm; in other words the interest is mostly on the structure, or topology, of the tree rather than on the exact values of the within / between groups distances. In addition the results of some hierarchical algorithms are not given in terms of distances; this is the case in particular with the original Ward's method, where the node levels represent variations of variance.

Kendall's tau (Kendall 1938) and Goodman-Kruskal's coefficient (Goodman and Kruskal, 1954) are both based on the ranks of the values being compared. Let $d(x_i, x_j)$ be the input distance between objects $x_i$ and $x_j$, and $u(x_i, x_j)$ the ultrametric distance between the same objects, resulting from a clustering algorithm. The $S^+$ index is the number of concordant pairs of distances, and $S^-$ is the number of discordant pairs ; two pairs of objects $(x_i, x_j)$ and $(x_k, x_l)$ constitute a "quadruple", they are said to be concordant if:

$$d(x_i, x_j) < d(x_k, x_l) \text{ and } u(x_i, x_j) < u(x_k, x_l)$$

they are said to be discordant if:

$$d(x_i, x_j) < d(x_k, x_l) \text{ and } u(x_i, x_j) > u(x_k, x_l)$$

Then the Goodman coefficient is:

$$GK = (S^+ - S^-) / (S^+ + S^-)$$

while the Kendall coefficient is:

$$\tau = (S^+ - S^-) / (N(N-1)/2)$$

where N is the number of distance pairs, that is $(n(n-1)/2)$ with n equal to the number of objects. These two coefficients differ by their denominator. Goodman-Kruskal denominator is the number of quadruples really taken into account (ties are not considered), while Kendall's denominator is equal to the number of all quadruples, including the possible ties. It seems not reasonable to take into account the tied pairs which may be numerous due to the ultrametric distances, induced by the hierarchy.

In addition the number of pairs really comparable may be much lower than in the case of a true correlation coefficient. For instance, in the dendrogram of Figure 1, pairs $(x_1, x_2)$ and $(x_1, x_4)$ may be compared : $(x_1, x_4) < (x_2, x_4)$ because the cluster including $x_1$ and $x_4$ is itself included in the cluster including both $x_2$ and $x_4$. On the other hand $(x_1, x_2)$ cannot be compared to $(x_2, x_4)$ because both pairs are included in the same set {$x_1$, $x_2$, $x_4$}. Again, no relation could be established between pairs $(x_1, x_2)$ and $(x_3, x_5)$ for the same reason. These remarks make the computation of Goodman-Kruskal coefficient a little more complicated than applying an existing software function.

## 6.3 Computing considerations

The computation of non-standard between cluster dissimilarities implies to preserve the initial distance matrix in the computer memory. Each step of the divisive process needs the examination of objects pairs as potential seeds for the dichotomy. Since there is no updating formula like in agglomerative algorithms, evaluating one dichotomy implies to recompute the splitting criterion for the two elements of the corresponding bipartition. Then this evaluation is of order $n^2$. The number of bipartitions is $O(n^2)$, therefore the complexity of one divisive step is $O(n^4)$. As the construction of the full binary hierarchy needs n – 1 steps, the overall complexity of the proposed divisive algorithms is $O(n^5)$. This involves a heavy computer task but is still possible for the moderate size of the target data.

### 6.4 Results and discussion

The experiment conducted according to the above conditions results in a table of 100 rows (random data sets) by 11 columns (algorithms). Any cell of this table includes the Goodman-Kruskal coefficient relative to one data set and one algorithm. The higher the coefficient the better is the corresponding algorithm, since this coefficient is akin to a correlation coefficient.

In Table 2 are gathered the average values of these coefficients over the 100 datasets. The algorithms are sorted by the average values of the Goodman-Kruskal coefficient in decreasing order. The best first two algorithms appear to be the "Silhouette based" and the "Dunn variant" methods. Next come the "Principal direction divisive" method and the "Divisive average link" method. But none of them can be declared as the best algorithm, since each of them, in turn, may show the best value of the Goodman-Kruskal coefficient, depending on the data at hand.

|  | Average | Std dev. |
|---|---|---|
| **Silhouette based** | 0.437 | 0.0387 |
| **Dunn's variant** | 0.428 | 0.0350 |
| **Principal direction (PDDP)** | 0.4181 | 0.0386 |
| **Average link divisive** | 0.4177 | 0.0403 |
| **Macnaughton-Smith *et al.*** | 0.398 | 0.0469 |
| **Ward Székély-Rizzo variant** | 0.395 | 0.0464 |
| **Ward's original formula** | 0.394 | 0.0430 |
| **Average link agglomerative** | 0.388 | 0.0371 |
| **Single link** | 0.368 | 0.0604 |
| **Dunn's original formula** | 0.366 | 0.0519 |
| **Complete link** | 0.282 | 0.0580 |

Table 2. Average values of the Goodman-Kruskal coefficient over the 100 random data sets.

In the lower part of this ranking appear three algorithms, namely those which are based on the single link and on the complete link, respectively, and the one based on Dunn's original formula. On average these three algorithms perform less well than the usual agglomerative average link method.

## 7 Conclusion

This study compared a set of 10 divisive hierarchical clustering algorithms taking as input dissimilarity datasets. The usual Average Link Agglomerative method was added as a basic reference. Five algorithms are based on the popular formulas used in pairwise aggregative procedures, namely the single link, the complete link and the average link methods and two versions of Ward's algorithm. The other 5 divisive algorithms are based on splitting criteria involving a ratio of between-group dissimilarities and within-group dissimilarities.

An important argument of the present work is that it is possible to separate the computation of the hierarchical node levels from the criterion used for splitting a cluster. The question of a readable dendrogram (without crossing branches over) is then solved by using the diameters of the clusters.

This does not hamper the internal evaluation of the results, that is to say the comparison of the hierarchy with the initial data, thanks to the Goodman-Kruskal correlation coefficient. Comparing the order relation induced by the successive inclusions of the clusters with the order relation associated to the input dissimilarities, this correlation coefficient provides us with an evaluation independant from the node levels, and concentrates on the shape of the dendrogram which is certainly the main interest of the user.

Applied to a sample of a hundred random datasets these principles allow for a ranking of the algorithms. The best ones are based on ratio-type splitting criteria: the Silhouette formula and a variant of Dunn's formula for partitions. At the lower end of this ranking appear the various Ward's procedures, Dunn's original formula, together with the complete link and single link based procedures, which dissuades to use them. This finding confirms some previous considerations on Ward's agglomerative hierarchical algorithm (Roux 2012, 2014).